\documentclass[preprint]{revtex4}

\newcommand{\be}{\begin{equation}}
\newcommand{\ee}{\end{equation}}
\newcommand{\bea}{\begin{eqnarray}}
\newcommand{\eea}{\end{eqnarray}}
\newcommand{\bml}{\begin{mathletters}}
\newcommand{\eml}{\end{mathletters}}

\usepackage{graphicx}
\usepackage{amsmath}
\usepackage{array}
\usepackage{bm}
\usepackage{amssymb}
\usepackage{amsfonts}
\usepackage{epsfig} 
\usepackage{epstopdf} 
\usepackage{color}

\begin{document}
\title{Gravitating bubbles of gluon plasma above deconfinement temperature}
\author{Yves \surname{Brihaye}$^1$}
\email[E-mail: ]{yves.brihaye@umons.ac.be}
\author{Fabien \surname{Buisseret}$^{2,3}$}
\email[E-mail: ]{fabien.buisseret@umons.ac.be}
\affiliation{$^1$ Service de Physique de l'Univers, Champs et Gravitation,
Universit\'{e} de Mons, UMONS  Research
Institute for Complex Systems, Place du Parc 20, 7000 Mons, Belgium\\ 
$^2$ Service de Physique Nucl\'{e}aire et Subnucl\'{e}aire,
Universit\'{e} de Mons, UMONS  Research
Institute for Complex Systems, Place du Parc 20, 7000 Mons, Belgium\\
$^{3}$ \quad CeREF, Chaussée de Binche 159, 7000 Mons, Belgium\\
}

\begin{abstract}
The equation of state of SU(3) Yang-Mills theory can be modelled by an effective $Z_3-$symmetric potential $V(\vert\phi\vert,\phi^3+\phi^{3*}, T)$ depending on the temperature $T$ and on a scalar field $\phi$ -- the averaged Polyakov loop. Allowing $\phi$ to be dynamical opens the way to the study of spatially localized classical configurations of the Polyakov loop. We first show that spherically symmetric static Q-balls exist in the range $(1-1.21)\times T_c$, $T_c$ being the deconfinement temperature. Then we argue that Q-holes solutions, if any are unphysical within our framework. Finally we couple the Polyakov-loop Lagrangian to Einstein gravity and show that spherically symmetric static boson stars exist in the same range of temperature. The Q-ball and boson star solutions we find can be interpreted as ``bubbles" of deconfined gluonic matter; their mean radius is always smaller than 10 fm.
\end{abstract}
\maketitle

\section{Introduction}

A fascinating feature of Yang-Mills theory is the existence of a deconfinement temperature, $T_c$, above which free color charges (free gluons) may propagate without being confined into color singlets \cite{Polyakov:1978vu,Susskind:1979up}. This deconfined phase can be thought as a ``gluon plasma", in analogy with the celebrated quark-gluon plasma experimentally created first at RHIC \cite{Arsene:2004fa,PHENIX:2018lia}, \textit{i.e.} the deconfined phase of full QCD.  

In Yang-Mills theory at nonzero temperature $T$, the Polyakov loop is defined as $L(T, \vec y)= P\, {\rm e}^{i\, g\int^{1/T}_0d\tau A_0(\tau, \vec y)}$, with $A_0$ the temporal component of the Yang-Mills field and $\vec y$ the spatial coordinates. $P$ is the path-ordering, $g$ is the strong coupling constant and units where $\hbar = c =k_B= 1$ are used. The Polyakov loop is such that $\left\langle L(T,\vec y) \right\rangle=0 $ $(\neq 0)$ when the theory is in a (de)confined phase~\cite{Susskind:1979up,Weiss:1981ev}. Since gauge transformations belonging to the center of the gauge group only cause $L(T,\vec y)$ to be multiplied by an overall factor, it has been conjectured that the confinement/deconfinement phase transition might be linked to the spontaneous breaking of a global symmetry related to the center of the considered gauge algebra. In the particular case of SU($N_c$), deconfinement might thus be driven by the breaking of a global Z$_{N_c}$ symmetry~\cite{Yaffe:1982qf,Svetitsky:1982gs}, with the following dimensionless order parameter
\begin{equation}\label{phi}
\phi=\frac{1}{N_c}{\rm Tr}_c L.
\end{equation}
The above color-averaged Polyakov loop $\phi$ is simply called Polyakov loop in the following. It is assumed to be a dynamical field. 

A nontrivial solution for $\phi(\vec y)$ vanishing at infinity would model a ``bubble" of deconfined gluonic matter. We have already shown the existence of such solutions at the deconfinement temperature in flat space-time and at large $N_c$ \cite{Brihaye:2012uw}. In the case $N_c=3$, according to strong-coupling expansion, the $Z_3$-symmetric Polyakov-loop potential depends on $\vert \phi\vert^2$, $\vert\phi\vert^4$ and $\phi^3+\phi^{*3}$ at the lowest-order \cite{Polonyi:1982wz,Gross:1983pk}. Nontrivial static configurations in $Z_3$-symmetric potentials have already been found in \cite{Gupta:2010pp,Jin:2015goa}. Most of the effort in the field has actually been devoted to study the temporal evolution of such solutions in close relation with thermalisation issues of experimentally observed quark-gluon-plasma \cite{Scavenius:2001pa,Fraga:2004hp,Gupta:2010pp,Gupta:2011ag,Mohapatra:2012ck}. Here we search for spherically symmetric static Q-ball solutions with a focus on conditions constraining their existence: temperature range, radial nodes, etc. Less standard solitons as Q-holes, never studied within a Polyakov-loop model, are also discussed \cite{Nugaev:2016wyt}.

Finally we couple our Polyakov-loop model to Einstein gravity. 
It is known that pure Yang-Mills theory in $3+1$ dimensions, 
when coupled to Einstein gravity, has ``particle-like" solutions 
which were first discovered in the seminal paper \cite{Bartnik:1988am}.
Within our approach the Yang-Mills degrees of freedom are replaced
by a complex scalar field whose associated Q-balls, when coupled to gravity,
are called boson-stars-- see namely the review \cite{Liebling:2012fv} 
for more recent references. To our knowledge, such a problem has never been addressed at finite temperature although researches devoted to ``QCD boson stars" (at $T=0$) are currently ongoing \cite{Kulshreshtha:2020scn}. We build gravitating solutions of static-boson-star-type, \textit{i.e.} spherically symmetric localized configurations of the Polyakov loop that lead to an asymptotically flat metric without singularity. 

\section{The Model}\label{secModel}
\subsection{Polyakov-loop Lagrangian}\label{PLL}

\begin{figure}[t]
\includegraphics[width=8cm]{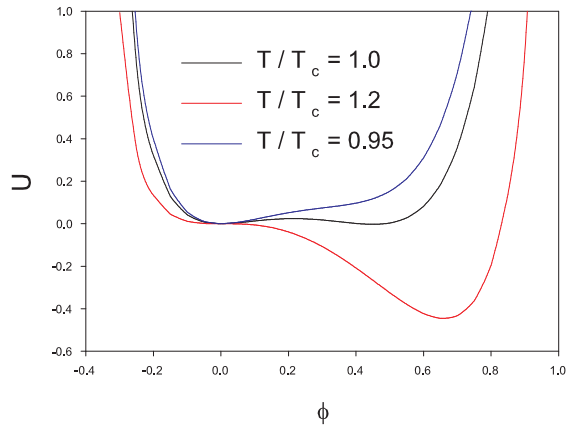}
\caption{\label{fig1}
Polyakov-loop potential $U(\phi,\phi^*,T)$ versus $\phi$ for various temperatures. $U$ is given by Eqs. (\ref{pot0}) and (\ref{params}), and the plot is restricted to $\phi\in\mathbb{R}$ for the sake of clarity. }
\end{figure}

Let us model SU(3) Yang-Mills theory at finite temperature by an effective Lagrangian based on the Polyakov loop (\ref{phi}) plus Z$_3-$symmetry. For the potential term $V_g$ we use that of Ref. \cite{Ratti:2006wg} which reads
\begin{equation}\label{pot1}
U(\phi,\phi^*,T)=\frac{V_g(\phi,\phi^*,T)}{T^4},
\end{equation}
with
\begin{equation}\label{pot0}
U(\phi,\phi^*,T)=-\frac{b_2(T)}{2} \vert\phi\vert^2 +b_4(T) \ln\left[1-6\vert\phi\vert^2+4 (\phi^3+\phi^{*3})-3 \vert\phi\vert^4 \right] .
\end{equation}
By definition $\vert\phi\vert<1$ classically and
\begin{equation}\label{params}
b_2(T)=3.51-2.47\, \frac{T_c}{T}+15.22\left(\frac{T_c}{T}\right)^2,\quad b_4(T)=-1.75 \left(\frac{T_c}{T}\right)^3.
\end{equation}
The above parameterisation leads to an optimal agreement with the equation of state of pure SU(3) Yang-Mills theory computed in lattice QCD \cite{Boyd:1996bx}. Potential (\ref{pot1}) is displayed in Fig. \ref{fig1} for the values (\ref{params}) of the parameters and for several temperatures. The change in minimum is clearly seen above and below $T_c$.  We notice that potential (\ref{pot1}) is only Z$_3$-symmetric and not U(1)-symmetric as it is often the case in matter Lagrangians based on a complex scalar field with typical potentials of the form  $|\phi|^6 - 2 |\phi|^4 + b |\phi|^2$ \cite{Volkov:2002aj}. A U(1)-symmetry can be recovered in the large-$N_c$ limit of Z$_{N_c}$-symmetric  potentials, see \cite{Buisseret:2011ms,Brihaye:2012uw}.

According to the suggestion of \textit{e.g.} Ref. \cite{Dumitru:2000in}, we allow the Polyakov loop to be a dynamical field and take a kinetic part of the form $N^2_c T^2 \partial_\mu \phi \partial^\mu \phi^* / \lambda$, which has both the correct energy dimensions and the expected $N_c$-scaling when the gauge group SU($N_c$) is chosen. $\lambda$ is the 't Hooft coupling. Minkowski metric has signature $(+---)$. In our SU(3) case, recalling that $\alpha_s=\lambda/(12\pi)$, one can write the matter Lagrangian as
\be
  {\cal L}_{phys} = \frac{3 T^2}{4\pi\alpha_s} \partial_\mu \phi \partial^\mu \phi^* -T^4 U(\phi,\phi^*,T),
\ee
where $\phi=\phi(y^\mu)$, $y^\mu$ are the spacetime coordinates. It is convenient to further define dimensionless variables $x^{\mu}$ related to the original (physical) ones by
\be
         y^{\mu} =  l_{phys} \ x^{\mu},\quad {\rm with}\quad  l_{phys}=\frac{\sqrt 3 }{T\sqrt{4\pi\alpha_s}} ,
\ee
so that the above Lagrangian can be replaced by the dimensionless one
\be \label{lag0}
          {\cal L}=\frac{ {\cal L}_{phys}}{T^4} = \partial_\mu \phi \partial^\mu \phi^* -U(\phi,\phi^*,T ) ,
\ee
where $\phi=\phi(x^\mu)$ and where $T$ is expressed in units of $T_c$.

It is worth estimating the physical length used in the model. First, a typical value for the deconfinement temperature in pure gauge QCD is $T_c=0.3$ GeV \cite{Sarkar:2010zza}. Second, a way to estimate $\alpha_s$ is to note that the short-range part of the static interaction between a quark and an antiquark scales as $-(4/3)\alpha_s/r$, at least from $T=0$ to $T_c$. Lattice studies, performed at $N_c=3$, favor $\alpha_s=0.2$ up to $T=T_c$ \cite{Kaczmarek:2005ui}, that is the value we retain here. We are then in position to estimate that,at $T=T_c$,
\be 
l_{phys}=3.6\ {\rm GeV}^{-1}=0.72\ {\rm fm}.
\ee
 
\subsection{Coupling to Einstein gravity} \label{EG}

The coupling of the above Lagrangian to gravity can be performed by minimally
coupling the scalar field to Einstein gravity: The action reads
\be \label{Sdef}
S=\int d^4x\sqrt{-g}\left(\frac{R}{\alpha}+{\cal L}\right),
\ee
with the effective coupling constant
\be
\alpha=16 \pi G_N l^2_{phys} T^4 .
\ee
The replacement of the partial derivatives by covariant ones in (\ref{lag0}) has to be performed. 

\section{Q-balls}\label{QB}
\subsection{Ansatz and existence conditions}
We begin by considering Lagrangian (\ref{lag0}) where $\phi^3$ and $\phi^{*3}$ are replaced by $|\phi|^3$ in order to recover the usual U(1)-symmetry needed to build Q-balls solutions. 

The classical equations of motion in flat space-time with potential $U(\vert\phi\vert,T)=-\frac{b_2(T)}{2} \vert\phi\vert^2 +b_4(T) \ln\left[1-6\vert\phi\vert^2+8 \vert\phi\vert^3-3 \vert\phi\vert^4 \right]$ read
\be\label{eom1}
     \partial_\mu \partial^\mu \phi = \partial_{\phi^*}  U=-\frac{b_2(T)}{2} \phi +6 b_4(T) \frac{-\phi+2\vert\phi\vert \phi- \vert\phi\vert^2\phi}{1-6 \vert\phi\vert^2-3 \vert\phi\vert^4+8\vert\phi\vert^3}
 \ee
plus the complex conjugated equation. We then perform the usual Q-ball ansatz on the scalar field~:
\be\label{ans1}
          \phi = \exp(i \omega t) \phi(r),
\ee
where $t=x^0$ and where $r=\sqrt{(x^{1})^2+(x^{2})^2+(x^{3})^2}$. The solutions we will build can be characterised by their mas $M$  and by a dimensionless conserved charge $Q$, respectively defined by 
\be \label{Mdef}
M =M_{phys} \int d^3 x \ T_{00}
\ee
with $M_{phys}=1/l_{phys}$ and
\be \label{Qdef}
Q =2 \omega \int d^3 x \ |\phi |^2 .
\ee
The temporal component of the energy-momentum tensor represents the energy density, given by
\be
T_{00}=\omega^2 |\phi |^2 + \vec\nabla \phi\cdot \vec\nabla \phi^* +U(|\phi |).
\ee
The conserved charge $Q$ finds its origin in the (artificially restored) U(1)-symmetry of the considered Lagrangian, leading to a conserved Noether current of the form  $J_{\mu} = i( \phi \partial_{\mu} \phi^*  - \phi^* \partial_{\mu} \phi)$, $Q$ being the the space integral of $J_0$. Axially symmetric solutions having  $k\neq 0$ are spinning Q-balls whose angular momentum $J$ is related to the charge $Q$ according to $J=k Q$ \cite{Volkov:2002aj}. Here we focus on non-spinning Q-balls.

We have studied the equations for generic values of $\omega$ although it is clear that only the solutions
corresponding to $\omega = 0$ are physically relevant for the potential under consideration in the Polyakov loop context: The original potential is Z$_3$-symmetric, not U(1). Note also that, if $\phi(r)$ is a real solution of the equations of motion, ${\rm e}^{\frac{i k\pi }{3}}$ with $k\in\mathbb{Z}$ is also a solution because of the system's symmetry.

The mass term of the potential plays a crucial role in the existence of the solutions. In a power expansion in $\vert\phi\vert$,
\be
            U(\vert\phi\vert, T)= m^2(T) |\phi|^2 + `{\rm higher\ order}" \ \ ,\ {\rm with }\ \ m^2(T) = - \frac{b_2(T)}{2} - 6 b_4(T).
\ee
General results on Q-balls \cite{Volkov:2002aj} state that the soliton exist
for  $\omega_{min} \leq \omega \leq \omega_{max}$ with
\be
        \omega_{min} = \min_{\vert\phi\vert} \frac{U(\vert\phi\vert, T)}{\vert\phi\vert^2}  \ \ , \ \ \omega_{max} = m(T).
\ee 
In particular, if the potential $U(\vert\phi\vert)$ is negative in some interval of values of $|\phi|$,
the value $\omega=0$ belongs to the spectrum of the boson star. This turns out to be
the case for $T > T_c$. The condition $m(T)^2 > 0$ also needs to be fulfilled;
in terms temperature, this corresponds to  $T/T_c > 1.21$.
As a consequence, the general properties of Q-balls solutions suggest that Q-ball solutions with zero frequency
will exist for $1 < T/T_c < 1.21$.  

\subsection{Numerical results}
A numerical resolution of the equations (\ref{eom1}) can now be performed. We use a collocation method for boundary-value ordinary differential equations, equipped with an adaptive mesh selection procedure \cite{colsys}. The regularity of the solution at the origin implies $\frac{d\phi}{dr}(r=0)=0$, the finiteness of the energy and the charge impose $\phi(\infty)=0$. These are the boundary conditions. 

\begin{figure}[t]
	\includegraphics[width=8cm]{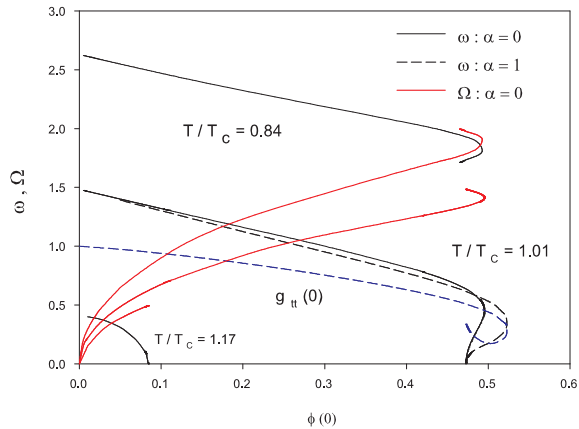}
	\caption{\label{fig4}
		Relation between $\omega$ and $\Omega \equiv \sqrt{m^2(T)- \omega^2}$ versus $\phi(0)$ for three values of $T/T_c$ in flat space-time ($\alpha=0$) (solid lines). The dotted lines represent
		$\omega$ and $g_{tt}(0)$ in the case $T/T_c = 1.01$ for gravitating solutions ($\alpha = 1$).} 
\end{figure}

\begin{figure}[t]
\includegraphics[width=8cm]{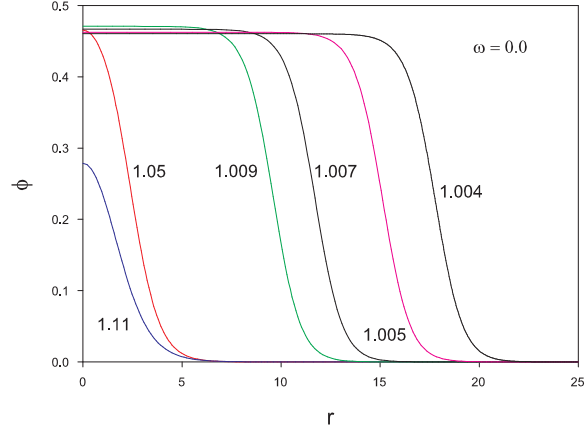}
\caption{\label{fig4b}
Profiles of $\phi(r)$ for several values of  $T/T_c$. Distances are in units of $l_{phys}$. }
\end{figure}

\begin{figure}[t]
\includegraphics[width=8cm]{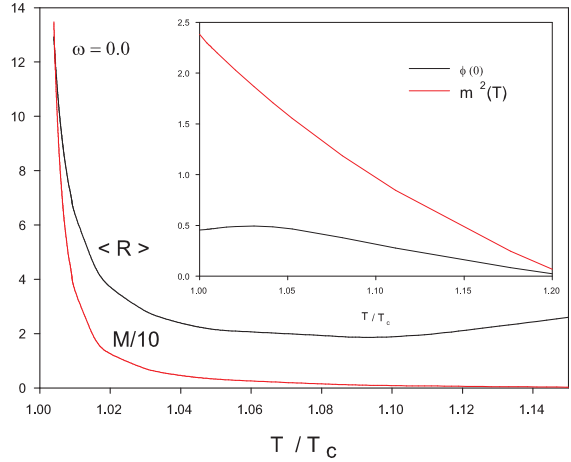}
\caption{\label{fig5}
Evolution of the mean radius $\langle R \rangle$ as a function of $T/T_c$, the insert contains $\phi(0)$ and the mass (\ref{Mdef}) of the scalar
field. Distances are in units of $l_{phys}$ and masses are in units of $M_{phys}$.}
\end{figure}

We present on Fig. \ref{fig4} the spectrum of the Q-balls for $T/T_c = 0.83,\, 1.01,\, 1.18$. It can be observed that no Q-ball solution with $\omega=0$ can be found below $T_c$: It is a nice feature of our model that it does not lead to solutions modelling deconfined matter below $T_c$. In the range $1 < T/T_c < 1.21$ suggested by the above analysis however, such solutions can be found. From now on, we concentrate on the latter $\omega=0$ solutions. Our results are summarized in Figs. \ref{fig4b} and \ref{fig5}.
Our numerical analysis shows that, in the limit $T/T_c \to 1.21$ the scalar function $\phi(r)$ approaches uniformly the
null function as expected by the existence criterion discussed before. The limit $T \to T_c$ reveals a peculiar  behaviour of the solitons: Their  mean radius and mass increase considerably. In this limit the scalar function $\phi(r)$ is closer and closer to a nonzero constant solution, leading to the observed increase in mass and mean radius.

 All the Q-balls we find have a mean radius smaller than 14$\times\, l_{phys}= 10$ fm and are lighter than $140\times\, M_{phys}=38.9$ GeV. Similar solutions were found in \cite{Gupta:2010pp} with a simpler, power-law, $Z_3-$symmetric potential of the form $\vert \phi\vert^2-a (\phi^3+\phi^{*3})+b\vert\phi\vert^4$. At $T=1.1\, T_c$ they find a soliton with a typical size of 1.5$-$2 fm while we find a Q-ball with mean radius 1.43 fm and mass 200 MeV at the same temperature. Other results are obtained in \cite{Gupta:2010pp} but in $2+1$ dimensions so they cannot be compared to ours. 

We have tried to construct radially excited solutions, \textit{i.e.} solutions where the radial function presents one or more
nodes, but so far, we can not find any. The absence of solutions presenting nodes for our model can be explained by the following argument. In general, the existence of node solutions is closely related to the shape of the effective potential
\be\label{veff}
            V_{eff}(\vert\phi\vert) = \frac{\omega^2}{2} \vert\phi\vert^2 - \frac{1}{2} U(\vert\phi\vert) \ .
\ee
Several conditions are necessary for node solutions to exist \cite{Volkov:2002aj,Kleihaus:2005me}: (i) $\phi = 0$ should be a local maximum of $V_{eff}$,
(ii) the effective potential should admit local minima for both signs of $\phi$.
It would be challenging to have a generic proof of the absence
of node-solutions with our potential but an inspection of the potential $U(\vert\phi\vert)$ quickly reveals that no local minimum exist for $\phi < 0$ when $\phi\in\mathbb{R}$ (see Fig. \ref{fig1}), so the condition (ii) cannot be fulfilled when $\omega=0$. 

The spectrum of the fundamental Q-ball appears quite different with the present potential than in more conventional U(1)-symmetric potentials. For example, one of us previously studied Q-balls with the SUSY-inspired potential $U_{SUSY}(\vert\phi\vert)\sim 1-\exp\left(-\frac{\vert\phi\vert^2}{\eta^2})\right)$, with $\eta\in\mathbb{R}^+_0$ \cite{Brihaye:2014gua}. In contrast to our potential, solutions can be constructed for arbitrarily large values of the central density $\phi(0)$ with the latter potential, and radially excited solitons can be obtained. 
We considered an effective potential consisting of  a linear superposition of our potential and the SUSY-potential. The latter is known to admit node solutions: $ U_{eff} = \cos(y)\, U_{SUSY} + \sin(y)\; U$ with $y \in [0, \pi/2] $. It turns out that when we progressively deform the SUSY-potential into our potential (say with a fixed value $\phi(0) < 1$) the zero-node solutions get continuously deformed and the frequency $\omega$ decreases with increasing the mixing parameter $y$. By contrast, for the one node solution, the $\omega$ quickly reaches $\omega = 1$ and the solution becomes oscillating.

\subsection{Symmetry breaking}

An obvious outlook is to include the matter sector of QCD in our approach. As discussed in \cite{Sannino:2005sk}, an immediate effect of quarks is the breaking of $Z_3-$symmetry in the Polyakov loop potential and the simplest way to mimic that symmetry breaking is to add a term proportional to $(\phi+\phi^*)$ or even $(\phi+\phi^*)^2$ to the potential. A more rigorous treatment of quark fields consists in resorting to a Polyakov-Nambu-Jona-Lasinio Lagrangian \cite{Fukushima:2003fw,Jin:2015goa}, in which the Polyakov-loop is coupled to the quark sector. Within mean field approximation for quarks, it has been show in \cite{Biswal:2019xju} that complex-valued solutions of Q-ball-type still exist with broken $Z_3$-symmetry. A question arising at this stage is therefore: May the $\omega=0$ Q-balls we constructed ``survive" to such a symmetry breaking ? 

We propose to perform the substitution $U\rightarrow  U_\beta=U+\beta\, (\phi+\phi^*)^2$ with $\beta$ a real constant parameter. This ansatz breaks the $Z_3$-symmetry while still allowing analytical calculations. In a power expansion in $\phi$ on the real axis, the $\beta$-term shifts the mass term: $m^2(T)\rightarrow m^2(T)+4\beta$. Graphical inspection of the modified potential shows that there is always an interval of $\phi$ values in which $U_\beta$ is negative above $T_c$ if $\beta<0$. Nevertheless a negative value of $\beta$ lowers $m^2$ and therefore lowers the maximal temperature at which Q-balls may be expected ($m^2$ is indeed a decreasing function of $T$). If $-0.5<\beta<0$, there always exists an interval of temperatures above $T_c$ for which the existence of Q-balls is guaranteed. We can thus safely assume that the solutions we find will not necessarily disappear in a more realistic theory including quarks.

\section{Q-holes}

Q-holes are configurations of the scalar field behaving like (\ref{ans1}) but such that $\vert\phi(r=0)\vert> 0$ and $\vert\phi(r\to\infty)\vert=\phi_c >\vert\phi(r=0)\vert$, with $\phi_c$ a local minimum of the potential under study \cite{Nugaev:2016wyt}. Are $\omega=0$ Q-holes solutions worth being constructed within the present approach?

A necessary condition for Q-holes to exist is that the second maximum of the effective potential (\ref{veff}) is lower than the maximum at the origin \cite{Nugaev:2019vru}. As illustrated in Fig. \ref{fig1}, it can only happen below $T_c$ in our model since  $V_{eff}(\vert\phi\vert) = - \frac{1}{2} U(\vert\phi\vert) $ at vanishing $\omega$. A configuration where the Polyakov loop is everywhere nonzero at a temperature below the deconfinement one is not physically relevant; although the problem is interesting from a technical point of view we have thus to discard Q-hole solutions in the present work.

\section{Boson stars}

The action (\ref{Sdef}) leads to the Einstein equation 
\be 
G_{\mu\nu}=\frac{\alpha}{2}T_{\mu\nu}
\ee
where the energy-momentum tensor is given by $T_{\mu\nu}=(D_\mu \phi)^*(D_\nu \phi)+(D_\mu \phi)(D_\nu \phi)^*-g_{\mu\nu}(D_\alpha \phi)^*(D^\alpha \phi)+g_{\mu\nu}{\cal L}$. The metric defines $ds^2=g_{\mu\nu}dx^\mu dx^\nu$ and $D_\mu$ is the covariant derivative. An estimation of this coupling constant $\alpha$ in the temperature range under study is $16 \pi G_N l^2_{phys} T_c^4=3.58\ 10^{-38}$. It is so small that 
no significant change of the solutions can be observed by numerical investigation. To appreciate more clearly the
influence of gravity on the system we will construct solutions with $\alpha =0.01$ and 1.

We search for boson-star solution. First, the ansatz (\ref{ans1}) and the boundary conditions are kept for the scalar field. Second, we choose a spherically symmetric ansatz for the metric:
\be
ds^2=-f(r) dt^2+\frac{l(r)}{f(r)}\left( dr^2+r^2\, d\theta^2+r^2 \sin^2\varphi\, d\varphi^2\right),
\ee
with the boundary conditions $f(r=+\infty)=l(r=+\infty)=1$ (asymptotically flat spacetime) and $f'(r=0)=l'(r=0)=0$ (no singularity at origin). The gravitational mass $M_G$ of these gravitating objects is defined as usual according to $f(r\to \infty) \sim 1 - \frac{2 M_G G_N}{r}$. The explicit equations involving $\phi$, $f$ and $l$ can be found in Appendix B of \cite{Kleihaus:2005me}; we do not recall them here for the sake of simplicity. The same numerical method as for Q-balls is used to construct boson-star solutions \cite{colsys}. 

\begin{figure}[t]
\includegraphics[width=8cm]{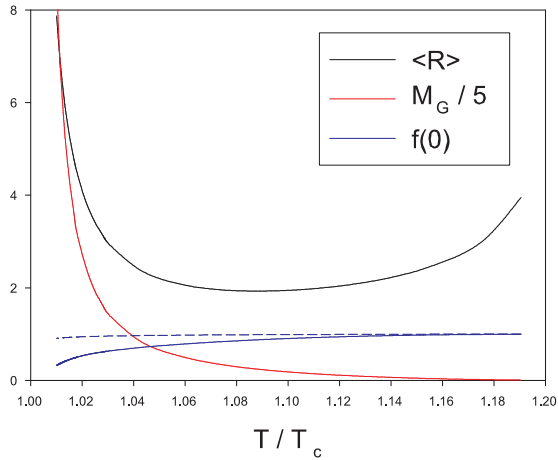}
\caption{\label{fig7}
Evolution of the mean radius $<R>$ (black line, in units of $l_{phys}$) 
and of the gravitational mass $M_G$ (red lines, in units of $M_{phys}$) 
as function of $T_c / T$ for $\omega = 0$ boson stars. The metric component $g_{00}= f(0)$ is represented by the solid (resp. dashed) blue lines for $\alpha=1$ (resp. $\alpha=0.1$). These values depend very weakly on $\alpha$ and the curves are mostly superimposed.}
\end{figure}

Even for such large  values as $\alpha=1$, our results indicate that gravitating solutions with $\omega=0$ still exist on roughly the same interval of $T/T_c$, see Fig. \ref{fig7}. However, the numerical analysis turns out to be tricky in the limit $T \to T_c$ likely because the 
local minimum of the potential disappears. Our results strongly suggest that
the gravitational mass and mean radius increase considerably
in the limit $T \to T_c$ as shown by Fig. \ref{fig7}, similarly as what is observed in the Q-ball case. As expected, the metric gets more deviated from the Minkowski 
metric in the central region of the soliton: 
for instance $g_{00} \ll 1$ (see the blue line of Fig. \ref{fig7}) 
and one can expect an essential singularity of the metric to be formed at $T_c$.

\begin{figure}[t]
\includegraphics[width=8cm]{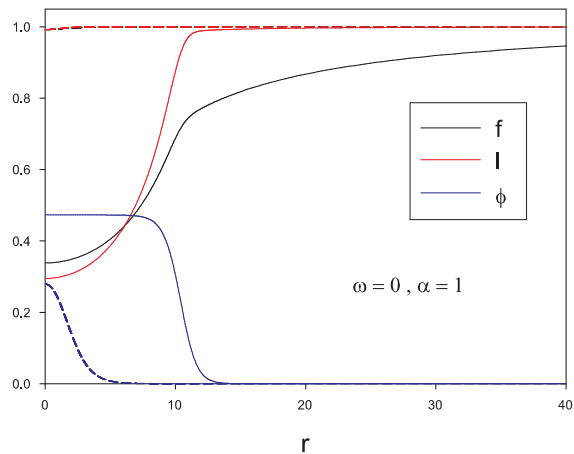}
\caption{\label{fig8}
 Profiles of the metric functions $f,l$ and of the scalar field $\phi$
for $\alpha=1 , \omega = 0$ and two values of the temperature : $T/T_c = 1.01$ (solid lines) $T/T_c = 1.11$ (dashed lines). The radial variable $r$ is in units of $l_{phys}$.
 }
\end{figure}

The profiles of the solution corresponding to $T/T_c = 1.01$ are presented in Fig. \ref{fig8} for $\alpha=1$
(solid lines). This plot clearly demonstrates that the soliton splits the space into two distinct regions:
an interior region where $\phi$ is practically constant and strongly curving space-time and a region with $\phi \sim 0$ where space-time is essentially Minkowski. These regions are separated by a ``wall" of the scalar field. The profile of a solution at an intermediate temperature $T/T_c = 1.11$ is also shown in Fig. \ref{fig8}; the same qualitative features are observed. The boson star finally presents different features for $T/T_c \to 1.21$, \textit{i.e.} the limit of vanishing $m(T)$. In this limit the scalar field approaches uniformly the null function and the Minkowski space-time is approached.

The existence of boson-star configurations for $\alpha=1$ implies the existence of such solutions for much smaller, ``realistic", values of the coupling constant around $T_c$, see Fig. \ref{fig7}. 

\section{Summary and outlook}\label{Conclu}

We have built Q-balls and boson stars from a model with a complex scalar field -- the Polyakov loop -- plus a temperature-dependent $Z_3$-symmetric potential mimicking Yang-Mills theory at finite temperature. We have shown that static Q-balls only exist between 1 and 1.21 $T_c$ with a mean radius smaller than 10 fm and that they cannot have radial nodes. The solutions we find are spherically symmetric and the scalar field is such that $\vert\phi(r=0)\vert\neq 0$ and $\vert\phi(r\to\infty)\vert= 0$; they can be interpreted as ``bubbles" of deconfined gluonic matter. We also showed that Q-holes solutions should be discarded from a physical point of view since they are solutions where the Polyakov loop is nonzero that can only exist below $T_c$ within our approach. Static boson stars exist in roughly the same temperature range as Q-balls. Their qualitative features are almost independent on the value of the matter-Einstein gravity coupling constant $\alpha$.

To our knowledge, it is the first time that boson stars are constructed from a Polyakov-loop potential such as (\ref{pot0}). Typical potentials used in boson-star-related studies are such that solutions exist for $0<\omega_{min} \leq \omega \leq \omega_{max}$, see \textit{i.e.} \cite{Brihaye:2014gua}. It is worth poiting out that the potential used here even allows the existence of static solutions with $\omega_{min}=0$.


Computation of the QCD equation of state in curved spacetime shows that the latter may affect the phase-diagram of the theory by increasing the splitting between the critical points for chiral and deconfinement transitions \cite{Sasagawa:2012mn}. We hope to present generalizations of our boson-star configurations to the case of a nontrivial quark field in a future work; they could shed new light on the interplay between confinement, chiral symmetry and gravity.

\bibliographystyle{unsrt}
\bibliography{BS}

\end{document}